\documentclass{article}

\usepackage{arxiv}

\usepackage[utf8]{inputenc} % allow utf-8 input
\usepackage[T1]{fontenc}    % use 8-bit T1 fonts
\usepackage{hyperref}       % hyperlinks
\usepackage{url}            % simple URL typesetting
\usepackage{booktabs}       % professional-quality tables
\usepackage{amsfonts}       % blackboard math symbols
\usepackage{nicefrac}       % compact symbols for 1/2, etc.
\usepackage{microtype}      % microtypography
\usepackage{lipsum}		% Can be removed after putting your text content
\usepackage{graphicx}
\usepackage{multirow}
\usepackage{natbib}
\usepackage{doi}
\usepackage{tabularx}
\usepackage{multicol}
\usepackage{xcolor}

\title{Using Psychological Characteristics of Situations for Social Situation Comprehension in Support Agents}

\author{ \href{https://orcid.org/0000-0002-1662-6652}{\includegraphics[scale=0.06]{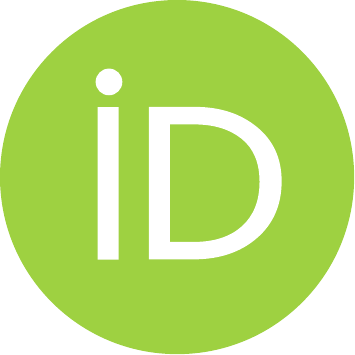}\hspace{1mm}Ilir Kola} \\
	Delft University of Technology\\
	The Netherlands \\
	\texttt{i.kola@tudelft.nl} \\
	\And
	\href{https://orcid.org/0000-0003-4780-7461}{\includegraphics[scale=0.06]{orcid.pdf}\hspace{1mm}Catholijn M.~Jonker} \\
	Delft University of Technology \& Leiden Institute for Advanced Computer Science \\
	The Netherlands \\
	\texttt{c.m.jonker@tudelft.nl} \\

	\And
	\href{https://orcid.org/0000-0001-9089-5271}{\includegraphics[scale=0.06]{orcid.pdf}\hspace{1mm}M. Birna van Riemsdijk} \\
	University of Twente\\
	The Netherlands \\
	\texttt{m.b.vanriemsdijk@utwente.nl} \\
}

\date{}

\renewcommand{\headeright}{}
\renewcommand{\undertitle}{}
\renewcommand{\shorttitle}{Psychological Characteristics of Situations for Social Situation Comprehension}

\hypersetup{
pdftitle={A template for the arxiv style},
pdfsubject={q-bio.NC, q-bio.QM},
pdfauthor={David S.~Hippocampus, Elias D.~Striatum},
pdfkeywords={First keyword, Second keyword, More},
}

\begin{document}
\maketitle

\begin{abstract}
    Support agents that help users in their daily lives need to take into account not only the user's characteristics, but also the social situation of the user. Existing work on including social context uses some type of situation cue as an input to information processing techniques in order to assess the expected behavior of the user. However, research shows that it is important to also determine the \emph{meaning} of a situation, a step which we refer to as social situation comprehension. We propose using psychological characteristics of situations, which have been proposed in social science for ascribing meaning to situations, as the basis for social situation comprehension. Using data from user studies, we evaluate this proposal from two perspectives. First, from a technical perspective, we show that psychological characteristics of situations can be used as input to predict the priority of social situations, and that psychological characteristics of situations can be predicted from the features of a social situation. Second, we investigate the role of the comprehension step in human-machine meaning making. We show that psychological characteristics can be successfully used as a basis for explanations given to users about the decisions of an agenda management personal assistant agent.
\end{abstract}

% keywords can be removed

\section{Introduction}\label{sec:introduction}

Artificial agents that support people in their daily lives -- such as personal assistants, health coaches, or habit formation support agents -- are becoming part of everyday lives (e.g. \cite{kepuska2018next, pinder2018digital}). Existing work on personal agents usually focuses on modelling personal characteristics of the user, such as their goals, emotional state, or personal values (e.g. \cite{kop2014personalized, fitzpatrick2017delivering, cranefield2017no}). However, research in social science shows that human behavior is not only shaped by a person's state and characteristics, but also by the situation they are in \citep{lewin1939field}. This suggests that in order to provide better aligned support, personal agents should  take the user's situation into account in determining which support to provide. 

In this paper we take a step towards addressing this challenge, with a specific focus on the \emph{social} dimension of situations. This is important because our daily situations often have a social nature: we spend time at work with colleagues, and free time with family and friends. Support agents thus need to account for the social dimension of situations, and how that affects the behavior of users. The need for enabling support agents to understand the social situation of the user has been acknowledged as an important open question in agent research \citep{tambe2008electric, van2015creating}. More broadly, the ability to assess and act in social situations has been proposed as the next challenge that intelligent machines should tackle \citep{grosz2012question}.

\subsection{Motivation}

Existing approaches (e.g., \cite{ajmeri2017arnor, dignum2014contextualized, kola2020predicting}) tackle this challenge by using some type of situation cues as input (e.g., actors, relationship characteristics, etc.), and using information processing techniques such as machine learning  or rule-based approaches to assess expected behavior. By going directly from social situation features to predicted or desired user behavior, the step of understanding the \textit{meaning} of the social situation from the point of view of the user is not performed explicitly. However, research in social psychology (e.g., \cite{edwards2005structure}) suggests that people determine how to behave in a situation by ascribing meaning to this situation, and using this interpretation to decide how to act. 

Inspired by this insight, \cite{kola2021ssa} propose that support agents should perform this step explicitly. They refer to this process as \textit{social situation comprehension}. Following research on situation awareness \citep{endsley1995toward}, they propose a three-level architecture where social situation comprehension is the middle level (Level 2) in between social situation perception (Level 1) and social situation projection (Level 3), as depicted in Figure~\ref{fig:ssa_arch}. The idea is that Level 2 information is derived from Level 1 information, i.e., social situation features, and Level 3 information about expected user behavior is in turn derived from Level 2 information.

A central question in realizing such a three-level architecture is in what `terms' the meaning of a situation should be described. In this paper we investigate whether \emph{psychological characteristics of situations}, a concept used in social psychology (e.g., \cite{parrigon2017caption, rauthmann2014situational, ziegler2014big}), can be used for this purpose of achieving social situation comprehension in support agents. The idea behind psychological characteristics of situations is that people view situations as real entities, and ascribe to them traits or characteristics in the same way they ascribe characteristics to other people. For instance, the situation `having a progress meeting with your supervisor' can have a high level of duty and intellect and a low level of deception and adversity. An important advantage of using psychological characteristics of situations is that they are general enough to model arbitrary daily life situations \citep{rauthmann2014situational}. 

Our goal is to explore whether incorporating information about the psychological characteristics of the user's situation would be beneficial for support agents. Support agents should make accurate suggestions that are trusted by the user. We investigate the use of psychological characteristics in support agents from these two perspectives. First, we study whether they can be used for predicting user behavior (Level 3 information), which is a basis for accurate suggestions. Second, we investigate whether they can provide meaningful reasons for explaining the suggestions of the support agent to the user, since research \citep{miller2019explanation} suggests that explainability of Artificial Intelligence (AI) systems is important for enhancing their understanding and in turn trustworthiness. 

\subsection{Use Case} \label{sec:use_case}
In this paper we take the example of a socially aware agenda management agent, inspired by the work of \cite{kola2020predicting}. Our goal is not to build a socially aware agenda management agent in itself, but this use case has characteristics that make it ideal for exploring the effects of incorporating psychological characteristics of situations. First of all, making accurate predictions on which to base its suggestions and giving insightful explanations is crucial for this agent, which is in line with aspects we aim to explore. Secondly, through this case we can study future situations for which the information is available beforehand. This way, we can focus on how the information can be processed to interpret the social situation and its effect on user behavior rather than having to deal with run-time situation perception, since that is beyond the purpose of our current work. Furthermore, such an agent facilitates conducting online user studies since it allows us to frame social situations as meetings, an easy concept to explain to participants. Lastly, the types of possible meetings can be arbitrary rather than about a specific domain, thus allowing us to explore a wide variety of social situations.

Providing support to the user regarding which meeting to attend can be seen as choice support. According to \cite{jameson2014choice}, in choice support the goal is to help the chooser (i.e., the user) make the choice in such a way that, from some relevant perspective, the chooser will be satisfied with the choice. \cite{jameson2014choice} present different choice patterns that people tend to follow and how technologies can support people in these choices: Access information and experience, Represent the choice situation, Combine and compute, Advise about processing, Design the domain and Evaluate on behalf of the chooser. The agenda management agent used throughout the paper gives suggestions to the users on which meetings to attend, thus following the `Evaluate on behalf of the chooser' choice support pattern.

\subsection{Research Questions and Hypothesis}
An important aspect of agenda management is dealing with scheduling conflicts where not all desired meetings can be attended. We develop predictive models that would allow such an agent to determine the priority level of each meeting, taking into account its social aspects. This is done via determining the situation profile of each meeting consisting of the psychological characteristics of the situation based on the DIAMONDS model \citep{rauthmann2014situational}. For example, dinner with a friend might be characterized by a low level of duty, but high level of positivity and sociality, while a meeting with a difficult colleague at work might be characterized by a high level of duty, high use of intellect and high level of adversity. This information is used to determine the priority level of each meeting, which is expected to correspond with the user behavior of choosing a high priority meeting in case of scheduling conflicts. The agent would make a suggestion to the user about which meeting to attend.

Based on this description, we formulate the following research hypothesis:

\begin{quote}
    \textbf{RH} - Using psychological characteristics of a social situation as input in a machine learning model leads to a more accurate prediction of the priority of the social situation than using social situation features as input.
\end{quote}

Collecting information about the psychological characteristics of each situation would be an intrusive task, therefore in the next research questions we explore whether we can automatically predict the psychological characteristics of a situation, and how useful would these predictions be:
\begin{itemize}
    \item \textbf{RQ1} - To what extent can we use machine learning techniques to predict the psychological characteristics of a social situation using social situation features as input?
    \item \textbf{RQ2} - To what extent can we use the predicted psychological characteristics from RQ1 as input in a machine learning model to predict the priority of a social situation?
\end{itemize}

Since we use explainable techniques for creating the predictive models, this also allows to determine which features were the most salient in determining the priority. These can be presented to the user as explanations. Following the previous example, if the two meetings are overlapping the predictive model might determine that the second meeting is more important and that the most salient feature is duty. In that case, the agent would tell the user `\textit{You should attend the second meeting since it involves a higher level of duty, and meetings with higher level of duty are usually prioritized}'. Through the following research questions we explore the perceived quality of such explanations:

\begin{itemize}
    \item \textbf{RQ3} - To what extent can social situation features and psychological characteristics of situations be used as a basis for explanations that are complete, satisfying, in line with how users reason, and persuasive?
    \item \textbf{RQ4} - When do people prefer psychological characteristics of situations in explanations compared to social situation features?
\end{itemize}

Our work has an exploratory nature, since the topic of incorporating psychological characteristics of situations in support agents is novel. For this reason, we do not always have a preconceived idea of the relation between variables to form hypotheses. Posing research questions allows us to explore and provide initial insights on the topic without being bound to specific expected outcomes. We assess these questions through two studies, one which addresses the predictive powers of psychological characteristics by creating machine learning models, and one which performs a user study to investigate the use of different kinds of explanations. The rest of the article is organized as follows: Section~\ref{sec:background} gives an overview of background concepts that we use throughout the paper. Section~\ref{sec:study1} introduces the first study, presents and discusses its results, and addresses \textbf{RH, RQ1} and \textbf{RQ2}. Section~\ref{sec:study2} introduces the second study, analyzes and discusses its results, and addresses \textbf{RQ3} and \textbf{RQ4}. Section~\ref{sec:conclusions} concludes the article.

\section{Background}\label{sec:background}
This section positions this paper in relation to existing work and offers an overview of background concepts that are used throughout the paper. In particular, we present the three-level social situation awareness architecture proposed in \cite{kola2021ssa} which forms the starting point for our work.

\subsection{Related Work}\label{sec:related_work}
The concept of sociality is broad, and so are its applications to artificial agents. The main directions involve agents being social with other artificial agents, and agents understanding human sociality. The agent technology research community has explored sociality from the point of view of artificial agents interacting with each other in multi-agent systems governed by structures such as norms, institutions and organizations (e.g., \cite{dignum2004model, fornara2007agent, maestro2020agent}). The other research direction explores the sociality of agents in relation to humans. This is researched from the perspective of agents interacting socially with people (e.g., \cite{davison2021words, elgarf2022and, valstar2016ask}), and agents modelling human sociality. An example of the latter is research on social signal processing, which focuses on using social cues such as body language to assess behavior \citep{vargas2021}. Other approaches more closely related to ours employ some type of social situation information as input, and process that information to assess expected user or agent behavior. In our work we take inspiration from the way in which they conceptualize social situations. The key difference is that we explicitly reason about the meaning of the social situation for the user.

\cite{dignum2014contextualized} propose using social practices \citep{reckwitz2002toward}. Social practices are seen as ways to act in context: once a practice is identified, people use that to determine what action to follow. For instance, the social practice `going to work' can incorporate the usual means of transport that can be used, timing constraints, weather and traffic conditions, etc. A social practice is identified using information from physical context, social context, activities, etc. Social context includes information about places and roles. Each social practice contains a concrete plan which makes the connection between the social context input and the behavior that needs to be manifested in that situation.

\cite{ajmeri2017arnor} also highlight the importance of modelling social context in personal agents. Social context includes information such as the place of the interaction or the social relationships between the people in the interaction (i.e., their role). In their approach, the agent includes the social information in the form of norms and sanctions that guide the agent's behavior. These norms and sanctions are formalized as rules in which the social context information serves as the antecedent and the behavior serves as the consequent: the agent exhibits a specific behavior only in presence of specific social context information.

Another approach on how to take into account the effects of social situations on user behavior is proposed in \cite{kola2020predicting}. They model social situations through a set of \textit{social situation features} seen from the point of view of the user. For instance, in a situation where a manager and an employee are meeting, the support agent of the employee would model this situation through features such as \textit{setting=work}, \textit{role of other person=manager}, \textit{hierarchy level=higher} and so on. Different from the previous approaches, in this work the relation between the social situation information and the expected behavior is learned rather than modelled explicitly. The authors show that it is possible to use these social situation features as input to a machine learning model to predict expected behavior such as the priority that people would assign to different social situations.

\subsection{Social Situation Awareness in Support Agents}
\label{sec:architecture}

\begin{figure}[]
    \centering
    \includegraphics[width=1\textwidth]{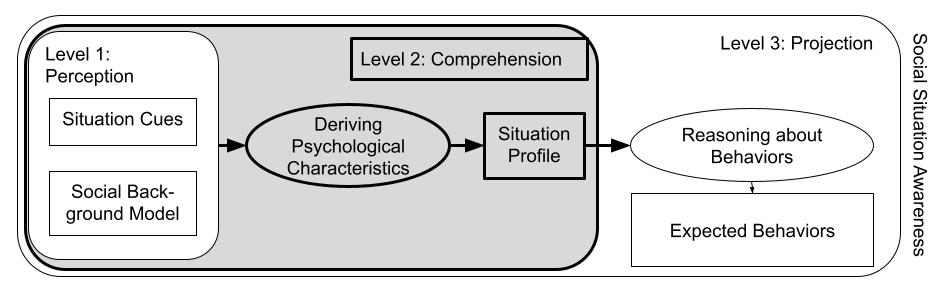}
    \caption{Simplified version of the three-level architecture for Social Situation Awareness proposed by \cite{kola2021ssa} (emphasis on Level 2 added by us).}
    \label{fig:ssa_arch}
\end{figure}

Our work builds on that of \cite{kola2021ssa}, who propose a three-level architecture for social situation awareness in support agents. They define social situation awareness as: \textit{``A support agent's ability to perceive the social elements of a situation, to comprehend their meaning, and to infer their effect on the behavior of the user"}. This definition instantiates Endsley's three-level model of situation awareness \citep{endsley1995toward}, yielding three corresponding levels of social situation awareness: social situation perception, social situation comprehension, and social situation projection. The resulting architecture is shown in Figure~\ref{fig:ssa_arch}. The focus of this paper is on the second level.

As can be seen from Figure~\ref{fig:ssa_arch}, one of the key parts of situation comprehension is the ability to use Level 1 information for deriving a situation profile at Level 2. A situation profile is intended to express the meaning of the situation for the user. Level 1 information concerns features that describe salient aspects of the social situation. This information can come via sensory input or interaction with the user. 

\cite{kola2019s, kola2020predicting} propose a set of features based on research from social sciences. They divide features into situation cues, namely \textit{setting, event frequency, initiator, help dynamic}, and social background features describing the social relation between the user and other people in the social situation, namely \textit{role, hierarchy level, contact frequency, geographical distance, years known, relationship quality, depth of acquaintance, formality level and shared interests.} In the rest of this paper we refer to these features as \textit{social situation features} or \textit{Level 1 information}.

The idea is that Level 1 information can be used to infer the meaning of the situation for the user, i.e,. Level 2 information. In this paper we investigate the use of psychological characteristics of situations to model Level 2. As proposed in social science research, psychological characteristics of situations  are used by people to ascribe meaning to a situation \citep{rauthmann2014situational}. People use these psychological characteristics to predict what will happen in a situation, and coordinate their behavior accordingly. There are five main taxonomies which provide a set of psychological characteristics to describe situations \cite{brown2015measuring, gerpott2018people, parrigon2017caption, rauthmann2014situational, ziegler2014big}, and in this work we use the psychological characteristics proposed in the DIAMONDS taxonomy \citep{rauthmann2014situational}. This taxonomy has several advantages. Firstly, it is intended to cover arbitrary situations, and it offers a validated scale for measuring psychological characteristics. Furthermore, it is shown that the psychological characteristics of a situation correlate both with the features of that situation and with the behavior people exhibit in that situation. The DIAMONDS taxonomy suggests that each situation can be described based on how characteristic each of the following concepts is: 
\begin{itemize}
    \item \textbf{Duty} - situations where a job has to be done, minor details are important, and rational thinking is called for;
    \item \textbf{Intellect} - situations that afford an opportunity to demonstrate intellectual capacity;
    \item \textbf{Adversity} - situations where you or someone else are (potentially) being criticized, blamed, or under threat;
    \item \textbf{Mating} - situations where potential romantic partners are present, and physical attractiveness is relevant;
    \item \textbf{pOsitivity} - playful and enjoyable situations, which are simple and clear-cut;
    \item \textbf{Negativity} - stressful, frustrating, and anxiety-inducing situations;
    \item \textbf{Deception} - situations where someone might be deceitful. These situations may cause feelings of hostility; 
    \item \textbf{Sociality} - situations where social interaction is possible, and close personal relationships are present or have the potential to develop.
\end{itemize}

We call such a description a situation profile. In the rest of this paper we also refer to the psychological characteristics of situations as \textit{Level 2 information}.

The idea is then that a situation profile can be used by a support agent to determine expected behaviors for the user (\emph{Level 3 information}), since research on the DIAMONDS model shows that there is a correlation between psychological characteristics of a situation and people's behavior in that situation. Information about expected behavior can in turn be used to determine how best to support the user.

\subsection{Explainable AI}

Following the definition of \cite{miller2019explanation}, when talking about explainable AI we refer to an agent revealing the underlying causes to its decision making processes. Early examples of such work can be found already more than forty years ago (e.g., \cite{scott1977explanation}). In the last five years, this field of research has received increasingly more attention\footnote{Google Scholar finds more than 22'000 publications in the time frame 2017-2022 for the search terms `explainable AI', which is more than the number of publications for the time frame 1955-2016.}. This is due to the increased availability of AI systems, as well as due to the emphasis on the importance of explainable AI coming from different governmental agencies \citep{goodman2017european, gunning2019darpa}. Different approaches have been proposed for explainable and interpretable AI (for an extensive survey, see \cite{mueller2019explanation}), and here we only provide a brief summary. Explanations can be global, i.e., explain the working of a system in general, and local, i.e., explain the reasons behind a specific decision or suggestion. Making the decisions of the agent explainable consists of three parts: the agent should be able to determine the internal processes that led to a certain suggestion, to generate an explanation based on them, and to present this explanation to the user \cite{neerincx2018using}. Different techniques have been proposed to determine the internal processes of so-called black box algorithms (for a survey, see \cite{guidotti2018survey}). When it comes to the content of explanations, research shows that shorter explanations explaining why a certain decision (rather than another decision) is made are preferred \citep{miller2019explanation, lim2009and}. Furthermore, \cite{ribera2019can} argue that explanations should be designed based on who the end user will be, and that explanations designed for lay users should be brief, use plain language, and should be evaluated via satisfaction questionnaires. We use these insights when designing the explanations for our user study.

\section{Study 1 - Predictive role of Psychological Characteristics}\label{sec:study1}

Through this study we evaluate our research hypothesis (\textbf{RH}), as well as \textbf{RQ1} and \textbf{RQ2}, as shown in Figure~\ref{fig:study1}.

\subsection{Method}\label{sec:study1_method}
In the first study we investigate to what extent psychological characteristics of situations can be used for predicting priority of meetings. Following the architecture in Figure~\ref{fig:ssa_arch}, a situation profile (Level 2) should be derived from Level 1 information, and it should be able to predict Level 3 information. In order to create corresponding predictive models, we use data from a user study that collects information at Level 1 (social situation features), Level 2 (psychological characteristics) and Level 3 (priority) for a range of meeting scenarios. 

\begin{figure}[]
    \centering
    \includegraphics[width=1\textwidth]{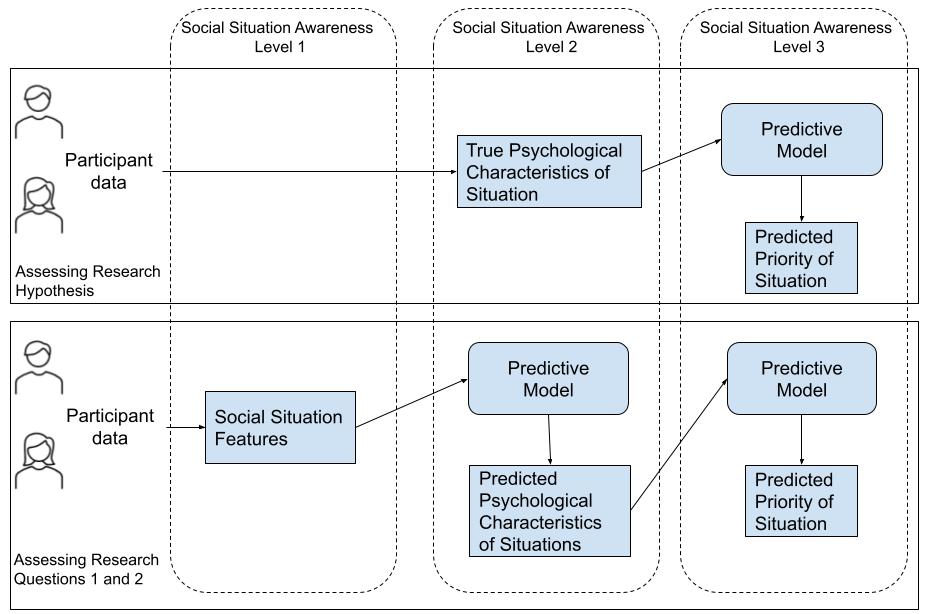}
    \caption{Conceptualization of Study 1, used to assess the research hypothesis (top part), and Research Questions 1 and 2 (bottom part)}
    \label{fig:study1}
\end{figure}

The data that we use for building the predictive models was collected through the experiment described in \cite{kola2020predicting}\footnote{The survey questions and the data can be found in: \url{https://doi.org/10.4121/16803889}}. The experiment was approved by the ethics committee of the university. Subjects were presented with meeting scenarios with people from their social circle (Level 1 information) and were asked to rate the psychological characteristics (Level 2 information) and priority of the meetings (Level 3 information). In their study, \cite{kola2020predicting} use only part of the collected dataset which involves the social situation features (see Section~\ref{sec:architecture}) and the priority of hypothetical social situations. In this work we also make use of information about the psychological characteristics of each of the hypothetical social situations. First, to assess whether priority could in principle be predicted from psychological characteristics of situations, we take the `true' Level 2 information as provided by our study participants, and create from this a predictive model for meeting priority (\textbf{RH}, top part of Figure~\ref{fig:study1}). While this allows to assess the possibility to predict Level 3 from Level 2, our agent would not have the `true' Level 2 information since it would be very cumbersome to ask users to provide this information for each meeting. This would not be the case for Level 1 information, since the social relationship features can be collected beforehand and tend to stay stable across situations. Thus, we want to investigate (see bottom part of Figure~\ref{fig:study1}) whether we can predict Level 2 information from Level 1 (\textbf{RQ1}), and in turn, use these predicted psychological characteristics as input to predict Level 3 information (\textbf{RQ2}) using the predictive model that was built to assess our \textbf{RH}. 

Data collection is a well-known obstacle when creating data-driven human decision predictive models. Using an experimental approach for collecting data is a good alternative when collecting data in the wild is not possible \citep{rosenfeld2018predicting}. Furthermore, such an experimental approach can allow for more flexibility in the type of data that is collected. In the data set that we are using, the experimental setup presents participants with hypothetical meeting situations involving real people from their social circle. These hypothetical meetings are highly diverse in terms of their priority level and relationship features of the participant and the other person, including situations work meetings with supervisors, family occasions, casual meetings with friends etc. Explicitly capturing every aspect that is involved in how the user assigns a priority level to the meeting is not possible in practice for such a wide variety of meetings. Therefore, our goal is to explore whether modelling psychological characteristics of the situations can provide a good approximation that leads to accurate predictions of the priority levels.

%In this section we briefly describe the method used to collect the data. For more details on the selection of concepts and methods we refer to the description in Kola et al. \cite{kola2020predicting}.

\subsubsection{Material}\label{sec:study1_material}
\emph{Social situation features} used in the study were based on literature from social science (see Section~\ref{sec:architecture} and \cite{kola2020predicting}). Specifically, the features used were: role of the other person, their hierarchy level, the quality of their relationship, the contact frequency, how long they have known each other, the geographical distance, the depth of acquaintance, the level of formality of the relationship, and the amount of shared interests.

\emph{Psychological characteristics of situations} were taken from the DIAMONDS taxonomy (see Section~\ref{sec:architecture}), namely Duty, Intellect, Adversity, Mating, Positivity, Negativity, Deception and Sociality.

\emph{Scenarios} used in this work represent social meeting settings that a user might encounter in their daily life. The scenarios had a hypothetical nature. Using hypothetical situations gives control over the types of situations subjects are presented with, ensuring a wide variety. To make these hypothetical situations more realistic, subjects were presented with activities that are common for people in their daily lives. Meeting situations were based on inputs from the users of a pre-study, and were formed as a combination of situation specific features (see Section~\ref{sec:architecture}): setting in which the meeting is taking place, frequency of meeting, initiator, and whether the user is expected to give or receive help (E.g. ``You have a weekly meeting with AB\footnote{For privacy reasons, users provided only the initials of people from their social circles.} where you expect to get feedback on a project that you are working on.''). In the situation descriptions, the setting was represented through typical activities that take places within that setting, to make the scenarios more concrete. For instance, the settings `work' and `casual' were represented by activities such as `having a meeting with the supervisor' and `going for dinner with a friend' respectively.

\subsubsection{Participants}
The study involved 278 subjects recruited through Prolific Academic\footnote{\hyperlink{}{https://www.prolific.co/}}, a crowd-sourcing platform where researchers can post studies and recruit participants who earn a monetary compensation for the time invested in conducting the study. 149 subjects were female, 127 were male, and 2 subjects selected the option `other'. The mean age was 36.2, with a standard deviation of 12.3. 

\subsubsection{Procedure}\label{sec:study1_procedure}
Subjects answered an online survey. First, participants were briefed about the purpose of the study. The goal of the study as conveyed to the participants was to collect information about the user's social relationships with different people from their social circle, as well as information about social situations involving the user and those people. Then they were presented with the two parts of the study. 

In the first part, subjects were asked to select five people from their social circle, and then were asked questions about their relationship with these people using the set of relationship specific features (see Section~\ref{sec:study1_material}). In the second part, subjects were presented with eight hypothetical social situations (see Section~\ref{sec:study1_material}), which were meeting scenarios between them and one of the people that they mentioned in the first part of the study (selected randomly). Subjects were asked what priority they would assign to each situation on a 7-point Likert scale (ranging from Very Low to Very High). 

Furthermore, subjects were asked about the psychological characteristics of each social situation using the dimensions proposed in the DIAMONDS taxonomy \citep{rauthmann2014situational} (see Section~\ref{sec:study1_material}). Subjects were presented with a description of each psychological characteristic, and they were asked ``How characteristic are each of the following concepts for this situation?". Subjects answered on a 6-point Likert scale, ranging from Very Uncharacteristic to Very Characteristic.

In total, the dataset consists of information about 1390 social relationships between the subjects and people from their social circle, and about the priority level and psychological characteristics of 2224 hypothetical social situations involving the subjects and one of these people.

\subsection{Results}\label{sec:study1_results}

The collected data is used to build predictive models\footnote{Code available in: \url{https://doi.org/10.4121/16803889}} which will be presented and evaluated in this section.

\subsubsection{Using Psychological Characteristics of Situations to Predict the Priority of Social Situations}\label{sec:study1_predict_priority}

The task of predicting the priority of social situations was previously explored by \cite{kola2020predicting}. In their work, they tested different learning algorithms that took as input the features of a social situation to predict the priority of that situation. If we refer to the social situation awareness architecture, this work takes as input Level 1 information and predicts Level 3 information. The best performing model was random forest, which led to a mean absolute error of 1.35, on a 7-points Likert scale. 

For this reason, we also employ a random forest model for predicting priority. In our case, the model takes as input the psychological characteristics of a social situation (Level 2), as obtained via the procedure described in the previous section, and predicts the priority of that social situation (as shown in Figure~\ref{fig:study1}, top). Specifically, we use the RandomForestRegressor implementation from the Scikit-learn package in Python. We split the data and randomly assign 80\% to the training set and 20\% to the test set. We perform parameter tuning by using cross validation on the training set. 

The results show that in our model, the mean absolute error is 0.98, which is a significant improvement (Wilcoxon Rank sum test, $p<0.05$) over the 1.35 mean absolute error reported by \cite{kola2020predicting}. This suggests that psychological characteristics of situations are a better predictor of the priority of social situations than social situation features, thus supporting our hypothesis (\textbf{RH}).

\subsubsection{Predicting the Psychological Characteristics of Social Situations}\label{sec:study1_predict_psych_char}

The social situation awareness architecture of \cite{kola2021ssa}, says that Level 2 information should be derived from Level 1 information. This is because having the agent ask the users about the psychological characteristics of each situation they encounter would be too invasive and time consuming. On the other hand, collecting Level 1 information can be done more efficiently, since the information about the social relationship can be collected in advance \citep{kola2021ssa}. For this reason, we investigate whether it is possible to predict the psychological characteristics of a social situation using as input social situation features (see Figure~\ref{fig:study1}, bottom). 

We evaluate the predictions of different regression algorithms: decision tree, XGBoost, Random Forest and Multi Layer Perceptor (MLP) using the scikit-learn library in Python. We train the models on 80\% of the data, and evaluate them on the remaining 20\%. We built 8 distinct models, where each model predicts one psychological characteristic, since this approach led to better accuracy than having one model that predicts all psychological characteristics at the same time. The model predicts a number from 1 to 6 (on a 6 point Likert scale, 1 being Very uncharacteristics, and 6 being Very characteristic), and the mean absolute errors are reported in Table~\ref{model_accuracies}. From the table (column `Random Forest') we can see that, for instance, the model is on average 1.17 off when predicting the level of Intellect for a social situation. This means that for instance, if the real value is 5 (i.e. Moderately characteristic), the model is expected to predict a value between 3.83 (i.e. Slightly characteristic) and 6 (i.e. Very characteristic). 

In order to assess how good these predictions are, we compare our models with a heuristic model that always predicts the mean of the psychological characteristics. The results are reported in Table \ref{model_accuracies} (column `Predict Mean'). We see that the random forest model significantly outperforms the heuristic predictor for all psychological characteristics apart from Adversity and Deception and always performs at least as well as the other predictive models. We use a heuristic model for comparison since this is the first benchmark result in predicting the psychological characteristics of a situation. Therefore we do not have an existing baseline to compare it with. Including heuristic baseline predictors is common practice for new machine learning tasks with no predetermined benchmarks (e.g. \cite{gu2018empirical}). \cite{kola2020predicting} also use heuristic predictors as a baseline for priority prediction, and the most accurate heuristic in that work is an algorithm that always predicts the mean priority. 

\newcolumntype{b}{X}
\newcolumntype{Y}{>{\hsize=.95\hsize}X}
\newcolumntype{s}{>{\centering\arraybackslash}Y}

\begin{table}[!h]
\caption{Mean Absolute Errors of the models in predicting the psychological characteristics of situations. Psychological characteristics marked with * represent statistically different results between the best performing model and the model that predicts the mean (Wilcoxon Rank sum test, $p<0.05$).}
\centering
\begin{tabular}{lccccc}
\hline
\begin{tabular}[c]{@{}l@{}}Psychological\\ Characteristic\end{tabular} & Decision Tree & XGBoost & Random Forest & MLP & Predict Mean \\ \hline
Duty*                                                                  & 1.66          & 1.36    & 1.34          & 1.38 & 1.55         \\
Intellect*                                                             & 1.48          & 1.21    & 1.17          & 1.23 & 1.3          \\
Adversity                                                              & 1.55          & 1.29    & 1.29          & 1.31 & 1.36         \\
Mating*                                                                & 0.92          & 0.87    & 0.85          & 0.93 & 1.03         \\
Positivity*                                                            & 1.44          & 1.18    & 1.14          & 1.17 & 1.26         \\
Negativity*                                                            & 1.51          & 1.25    & 1.25          & 1.39 & 1.37         \\
Deception                                                              & 1.14          & 1.04    & 1.04          & 1.09 & 1.09         \\
Sociality*                                                             & 1.42          & 1.06    & 1.02          & 1.03 & 1.13         \\ \hline
\end{tabular}
\label{model_accuracies}
\end{table}

In the next section we evaluate whether these predictions are sufficiently accurate to be used as an intermediate step for predicting priority of social situations. This allows the evaluation of the usefulness this predictive model as part of the bigger social situation awareness architecture. 
%of these predicted values. To do that, in the next section we assess their ability to predict the priority of social situations.

\subsubsection{Predicting Priority through Predicted Psychological Characteristics} \label{sec:study1_predict_priority_predicted_char}
To assess the usefulness of these predicted values for predicting the priority of social situations, we predict  priority by using as input not the `true' psychological characteristics of the situation as reported by the participants in the data collection experiment, but the predicted ones (Figure~\ref{fig:study1}, bottom). To do this, we use the model trained in Section~\ref{sec:study1_predict_priority}, and feed as input the predicted psychological characteristics from the Random Forest model in Section~\ref{sec:study1_predict_psych_char}. 

The model achieves a mean absolute error of 1.37 (Table~\ref{priority_accuracies}). As expected, there is a drop compared to the 0.98 error that we got using as input the true psychological characteristics. Nevertheless, we notice that the prediction error is not significantly worse than the results reported in \cite{kola2020predicting}, despite using predicted values as input (\textbf{RQ2}). This confirms the predictive potential of the psychological characteristics of situations. However, it also suggests the need for more research towards predicting these psychological characteristics more accurately, since that would lead to an overall better prediction of the priority of social situations.

\newcolumntype{b}{X}
\newcolumntype{Y}{>{\hsize=.32\hsize}X}
\newcolumntype{s}{>{\centering\arraybackslash}Y}

\begin{table}[!h]
\caption{Mean Absolute Errors of the models in predicting the priority of social situations when using different inputs. Results marked with * are significantly different from the others (Wilcoxon Rank sum test, $p<0.05$).}
\centering
\begin{tabularx}{0.95\textwidth}{bs}
\hline
Model input                     & \begin{tabular}[c]{@{}c@{}}Mean Absolute Error \\ in Priority Prediction\end{tabular} \\ \hline
Social situation features \citep{kola2020predicting}         & 1.35                                                                                             \\
True psychological characteristics of situations  & 0.98*                                                                                                                     \\
Predicted psychological characteristics of situations                                                             & 1.37                              
                                                          \\ \hline
\end{tabularx}
\label{priority_accuracies}
\end{table}

\section{Study 2 - Evaluating Explanations}\label{sec:study2}

In this section we present the setup of the user study we performed to evaluate explanations given by a hypothetical personal assistant agent about why they suggest attending a specific meeting, based on Level 1 and Level 2 information (\textbf{RQ3} and \textbf{RQ4}). 

In this study\footnote{The survey questions and the data can be accessed in: \url{https://doi.org/10.4121/16803889}}, subjects were presented with pairs of social situations (in this case, meetings), and suggestions from a personal assistant agent regarding which meeting to attend, followed by an explanation that included as a reason either Level 1 or Level 2 information. Subjects were asked to evaluate these explanations (Figure~\ref{fig:study2}). The results of this study are presented in the next section.

\subsection{Design Choices and Material}\label{sec:study2_design_choices}
In this section we present the choices we made in the design of the experiment, and the resulting material used for conducting it.

\subsubsection{Simplifications} \label{sec:study2_simplifications}
This study falls under the human grounded evaluation category proposed by \cite{doshi2017towards}: a study with real humans, and a simplified task. The first simplification we made had to do with the fact that subjects were presented with hypothetical scenarios and explanations. This simplification was necessary since we do not yet have a fully fledged support agent ready to use and be tested in practice. Since the proposed scenarios were provided by us rather than by the participants themselves, this comes with the risk that participants may not actually encounter that particular situation themselves in their own lives directly (e.g., some scenarios refer to meetings with work colleagues, however the participant might not be employed). For this reason, in this study we opted for a third-person perspective, i.e., asking participants to imagine how another user might evaluate the explanation if they were to encounter that scenario. Moreover, using existing scenarios allowed us to balance which psychological characteristics were used, which was important for investigating whether people hold different preferences for different characteristics. The second simplification had to do with the fact that the explanations were not formed using a specific explainable AI method, but designed by the researchers based on insights from our predictive models in Section~\ref{sec:study1_results}. 

In order to make the hypothetical setting as realistic as possible, scenarios were retrieved from the the data collected by \cite{kola2020grouping}. In that study, subjects described social situations from their lives, and answered questions about the psychological characteristics of those situations (Level 2). However, the dataset did not include annotated Level 1 information, which is needed to form the explanations based on this type of information. To perform the annotation, we used information that is available in the description of the situations. For instance, if the description says `I am meeting my boss to discuss the project', we infer that the role of the other person is \textit{supervisor}, the hierarchy level is \textit{higher} and the setting is \textit{work}, and consider the information that is not available in the description to be equal across situations. Using only explicit information available in the description to infer Level 1 information allows this procedure to be unambiguous. At this point, we have a dataset with situations described by people, annotated in terms of their social situation features and psychological characteristics which will be used to form the explanations.

\subsubsection{Selecting which information is included in explanations}\label{sec:study2_explanation_content}

For an explanation to be realistic, it needs to be based on information that contributed to the suggestion of the agent. In order to find the Level 1 and Level 2 information that is more likely to have contributed to the priority prediction, we identified the features that have the highest weight when predicting the priority of social situations using the TreeExplainer method of the SHAP package \citep{lundberg2017unified}. For Level 1, these features were setting, help dynamic, role, relationship quality, age difference, and shared interests. For Level 2, these features were duty, intellect, positivity and negativity. We assume that the best explanation can be found in this pool of features, since they are the best predictors of priority. 

\subsubsection{Selecting scenarios}\label{sec:study2_selecting_scenarios}

We want users to evaluate the type of information included in the explanations, rather than evaluate whether the agent selected the right feature to include in the explanation. To facilitate this, we formed pairs of scenarios in such a way that both meetings have a set of common situation features/psychological characteristics and a single differing one, which would then be used in the explanation. This was done using the following procedure:

\begin{itemize}
    \item \textit{Level 1 - } Each meeting is annotated with a set of social situation features. To form pairs, we selected scenarios that have the same amount of information in terms of social relationship features (i.e., same number of social situation features known), and that differ in only one social relationship feature.
    \item \textit{Level 2 - } Each meeting is annotated in terms of its psychological characteristics, rated on a scale from 1 (very uncharacteristic of the situation) to 7 (very characteristic of the situation). We consider psychological characteristics with a score higher than 4 to have a \textit{high relevance} in the situation, and those with a score lower than 4 to have \textit{low relevance}. To form pairs, we selected scenarios that have a similar level of relevance (i.e., either high or low) for all psychological characteristics except for one, which has a differing level of relevance.
\end{itemize}

In total we formed eight pairs of scenarios, where the differing social relationship features were setting, help dynamic, role, relationship quality, age difference, and shared interests. The differing psychological characteristics were duty, intellect, positivity and negativity (two pairs for each). For instance, one of the pairs was:\\

\textit{Meeting 1 - }Alice has planned to meet a colleague because they want to update each other about their work.

\textit{Meeting 2 - }Alice has planned to meet another colleague because the colleague needs her help to solve a work task.\\

In this case the differing social relationship feature was the help dynamic\footnote{The feature \textit{help dynamic} can take the values \textit{giving help, receiving help, neither giving nor receiving help.}}, which was \textit{neither giving nor receiving help} for the first meeting and \textit{giving help} in the second (as inferred from the scenario descriptions), whereas the differing psychological characteristic is the level of duty, which was higher in the second meeting (as annotated by the subjects who proposed these scenarios).

\subsubsection{Selecting Agent Suggestions}\label{sec:study2_agent_suggestions}

To determine which meeting the agent should suggest the user to attend, we used a heuristic procedure based on the prediction models from Section~\ref{sec:study1_results}. Through the TreeExplainer method \citep{lundberg2017unified} we determined whether each differing feature contributes to a higher or a lower priority level. Since meetings differ in one feature (for each of Level 1 and Level 2), that feature is used as the tie breaker to determine which scenario should have higher priority. Scenarios were selected in such a way that the agent would make the same suggestion regardless whether it uses Level 1 information or Level 2 information for the prediction. This was done to minimize the effect that the agent suggestion has on the evaluation that the subjects give about the explanations.
For the aforementioned pair, Meeting 2 has a higher priority because, based on the prediction models:
\begin{itemize}
    \item Meetings where someone is expected to give help have a higher priority (Level 1 information);
    \item Meetings with a higher level of duty have a higher priority (Level 2 information).
\end{itemize}

\subsubsection{Selecting explanations}\label{sec:study2_selecting_explanations}

To form the explanations, we followed insights from research on Explainable AI which suggests using shorter explanations that have a comparative nature \citep{miller2019explanation, van2021evaluating}. For this reason, explanations include only the differing feature between the meetings (one for each explanation), and are phrased as comparisons between the available choices. For the previously introduced pair of scenarios, the explanations would be:\\

\textit{Explanation based on Level 1 information - }Alice should attend Meeting 2 because she is expected to give help, while in Meeting 1 she isn't, and meetings where one is expected to give help are usually prioritized.

\textit{Explanation based on Level 2 information - }Alice should attend Meeting 2 because because it involves a higher level of duty, which means she is counted on to do something, and meetings involving a higher level of duty are usually prioritized.

\subsection{Measurement}\label{sec:study2_measurement}
In order to evaluate how good the explanations are, we first need to decide on a set of criteria based on which they can be evaluated. \cite{vasilyeva2015goals} suggest that the goal of the explainer is key in how the explanations are evaluated. Different goals of explainable systems identified in the literature are transparency, scrutability, trust, persuasiveness, effectiveness, education, satisfaction, efficiency and debugging \citep{chromik2020taxonomy, tintarev2012evaluating, wang2019designing}. In our setting, the objective of the personal assistant agent is to justify its suggestions so the user can decide to accept them or not. Therefore, its main goal is to offer clear and understandable explanations for the reasons behind the suggestion, which relate to the goals transparency and satisfaction. Furthermore, we want to assess the persuasive power of the explanations. 

To assess how clear the explanations are, we use an adapted version of the explanation satisfaction scale \citep{hoffman2018metrics}. From the scale, we use the following statements:
\begin{itemize}
    \item The explanation of [...] is \textit{satisfying};
    \item The explanation of [...] has \textit{sufficient detail};
    \item The explanation of [...] seems \textit{complete};
\end{itemize}
We do not include the items of the scale that refer to accuracy, trust, usefulness to goals and whether the explanation tells the user how to use the system, since these items are not related to the goals of the envisioned support agent. 

To further inquire about the clarity and understandability of the explanations, we add the following statement:

\begin{itemize}
    \item The explanation of [...] is in line with what you consider when making similar decisions;
\end{itemize} 

This is done because we expect that being presented with information which is similar to what they consider when making similar decisions would make the explanations more understandable for the user.

Lastly, another goal of the agent is persuasiveness, which means how likely are the explanations to convince the user to follow the suggestion. This was captured through the following question:

\begin{itemize}
    \item The explanation of [...] is likely to convince Alice to accept the suggestion.
\end{itemize}

These items were rated on 5-points scales which were different for each experimental setting, as specified in Section~\ref{sec:study2_procedure_2.1} and Section~\ref{sec:study2_procedure_2.2}.

\subsection{Participants}
In total, we recruited 290 subjects through the crowd-sourcing platform Prolific Academic. Participation was open to members that had listed English as their first language. Every subject was compensated for the time they spent completing the study, as per the guidelines of the platform. The study consisted of two experiments. For the first experiment we recruited 100 subjects. Of these, 55 were female, and 45 were male, with a mean age of 31.1 and a standard deviation of 11.8. For the second experiment we recruited 190 subjects. Of these, 108 were female, 80 were male, 1 selected the option `other', and 1 selected the option `prefer not to say'. They had a mean age of 29.98 with a standard deviation of 10.28.

\subsection{Procedure}\label{sec:study2_procedure}

\begin{figure}[!h]
    \centering
    \includegraphics[width=1\textwidth]{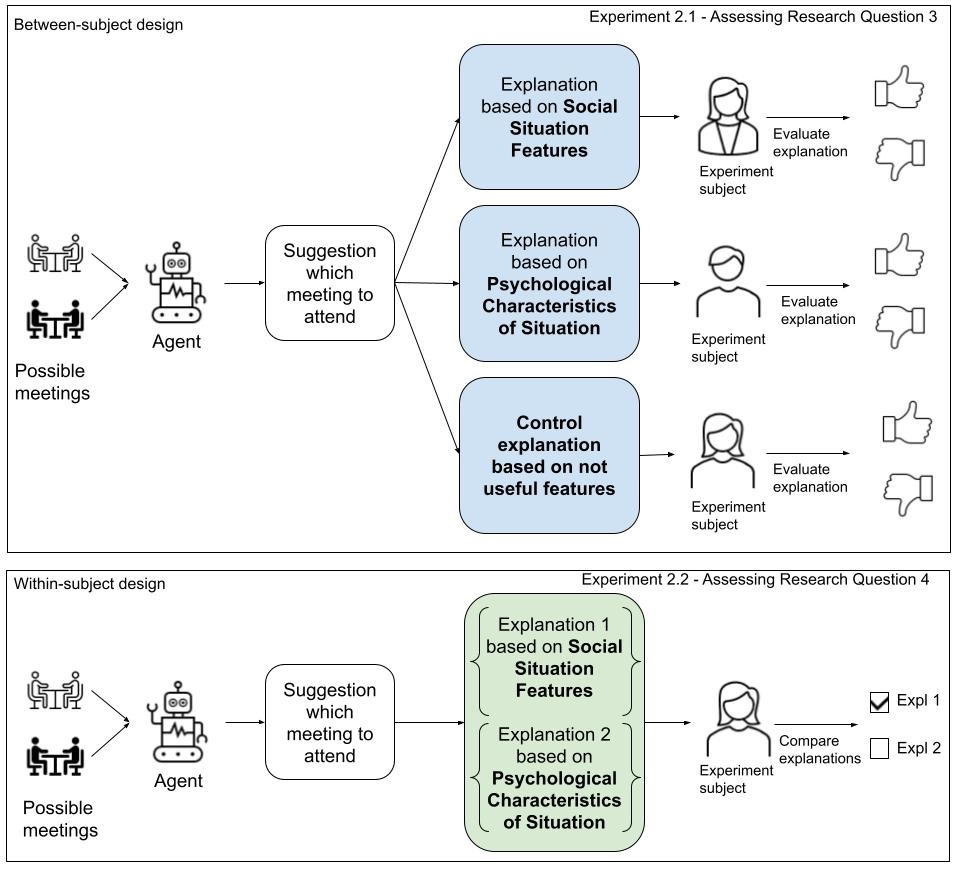}
    \caption{Conceptualization of Study 2, used to assess Research Questions 3 (top part) and 4 (bottom part).}
    \label{fig:study2}
\end{figure}

In this section we introduce the procedure that was used for this study. The study consisted of two experiments. In the first experiment (between-subject design, \textbf{RQ3}, top part of Figure~\ref{fig:study2}), participants are shown either an explanation based on social situation features (Level 1 information), psychological characteristics of the situation (Level 2 information), or a control explanation based on features that were considered not useful. In the second experiment (within-subject design, \textbf{RQ4}, bottom part of Figure~\ref{fig:study2}), we show participants both Level 1 and Level 2 explanations for a specific suggestion by the agent, and ask them to \emph{compare} these explanations and indicate which one they prefer. Both experiments were conducted as online surveys, and the subjects were recruited through the crowd-sourcing platform Prolific Academic. The study received the approval of the ethics committee of the university. The experimental procedure was similar in both experiments:
\begin{itemize}
    \item \textit{Introduction -} Subjects were informed about the study and were presented with the consent form.
    \item \textit{Demographics -} Subjects were asked about their age and gender to check whether the population sample was sufficiently broad.
    \item \textit{Case-study -} Subjects were introduced to Alice, a hypothetical user of the socially aware personal assistant agent. Subjects were told that during a specific week Alice is particularly busy, so the agent makes suggestions which meetings she should attend and which ones she should cancel. 
    \item \textit{Scenarios -} Subjects were presented with a pair of meeting scenarios, and they were asked which meeting they would suggest Alice to attend. This was asked to control for biases that they would have regarding the agent's suggestions, in case their own opinion differed from that of the agent. Furthermore, in an open question they were asked about the reasons behind this suggestion. This was asked to get more insights into the reasoning process of subjects in such situations. In total subjects were presented with four pairs of scenarios.
    \item \textit{Evaluation of explanations -} Subjects that made suggestions in line with the agent were presented with the full questionnaire which included all measures from Section~\ref{sec:study2_measurement}. Subjects that made suggestions that were different from what the agent would suggest were presented with a question regarding the persuasiveness of the different explanations (namely: ``The explanation offers convincing arguments''). This was done to take into account biases: We expect that subjects that do not agree with the agent suggestion would be implicitly evaluating the suggestion rather than its explanation. 
\end{itemize}
In the next subsections we present the specifics of each experiment.

\subsubsection{Experiment 2.1}\label{sec:study2_procedure_2.1}
This part of the study had a between-subjects design. Subjects were presented either with explanations based on Level 1 information, Level 2 information, or they were part of the control group, which we added to serve as a baseline. In related work (e.g., \cite{van2021evaluating}), control groups normally do not include an explanation, since the goal is usually to evaluate the impact of the explanation in the overall quality of the suggestion. However, in our setting that would be obsolete since the questions specifically refer to explanations. For this reason, in the control group subjects were presented with explanations that included information that could in principle be useful for determining the priority of meetings, but did not make sense for those specific scenarios. Explanations in the control group included information such as weather, geographical location or time. For instance, an explanation was ``Alice should attend the first meeting because it is spring". 

This design presents subjects with only one type of explanation, so the evaluation is absolute rather than relative to the other explanation types. This allows us to answer \textbf{RQ3}: to what extent can social situation features and psychological characteristics of situations be used as a basis for explanations?  

The aforementioned measurements were presented as statements such as ``The explanation provided about the reasons why the agent suggests Meeting 2 is satisfying". Subjects could answer on a 5-point Likert scale, ranging from Strongly disagree to Strongly agree.

\subsubsection{Experiment 2.2}\label{sec:study2_procedure_2.2}
This part of the study had a comparative within-subject design. This design presents subjects with two explanations for each pair of scenarios: one based on Level 1 information, and one based on Level 2 information. Through this setting, we address \textbf{RQ4}: when do people prefer one type of explanation versus the other? The measurements were framed as comparisons, for instance ``Which explanation do you consider more satisfying?". Subjects could answer `Significantly more Explanation A', `Slightly more Explanation A', `Both equally', `Slightly more Explanation B' and `Significantly more Explanation B'.

\subsection{Results and discussion} \label{sec:study2_results}

In this section we present the quantitative results of the two user studies described above, and we analyze the answers to the open question.

\subsubsection{Experiment 2.1} \label{sec:study2_results2.1}

\begin{figure}[]
    \centering
    \includegraphics[width=1\textwidth]{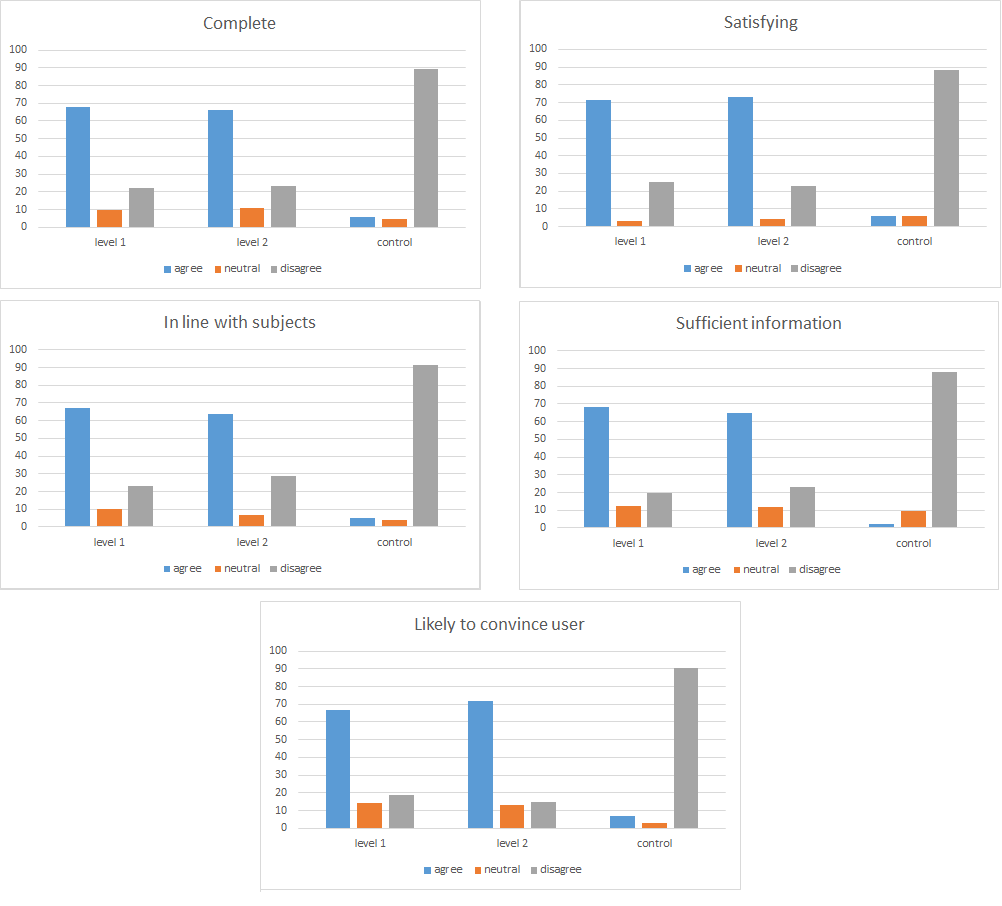}
    \caption{Answer distributions for the different measurements. The $x$ axis represents the answer options for each of the levels. `Strongly agree' and `Somewhat agree' were grouped together as `agree', and `Strongly disagree' and `Somewhat disagree' were grouped together as `disagree'. The $y$ axis shows the percentage of subjects that gave a specific answer.}
    \label{fig:between_subject_results}
\end{figure}

Each of the subjects was presented with four pairs of scenarios, which means 400 pairs of scenarios were shown to subjects across the different conditions (128 pairs in the Level 1 group, 140 pairs in the Level 2 group, and 132 pairs in the control group). In 73\% of the total cases, subjects would suggest Alice to attend the same meeting that the agent would suggest. Figure~\ref{fig:between_subject_results} presents the subjects' answers for each of the measurements regarding the explanation provided by the agent. This applies to the subjects whose suggestions were in line with the suggestions of the agent.

The majority of the subjects considered the explanations based on Level 1 or Level 2 information to be complete, satisfying, in line with how the subjects reason, likely to convince the user, and having sufficient information. While explanations based on Level 1 or Level 2 information were thus considered positively, on the other hand, subjects strongly disliked the explanations offered in the control setting. This confirms that the positive effect was not just due to the presence of an explanation as such, since subjects do not give a positive evaluation to an explanation which does not apply to the suggestion.

\begin{figure}[]
    \centering
    \includegraphics[scale=0.75]{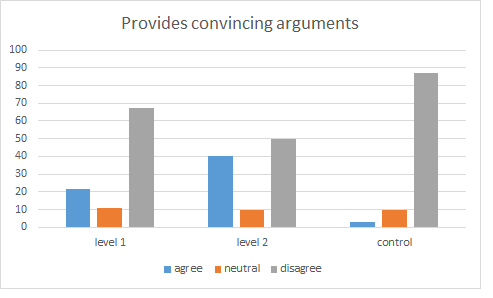}
    \caption{Answer distribution for the subjects who would make a suggestion different from the agent's.}
    \label{fig:disagreement}
\end{figure}

The answers of the subjects whose suggestions were not in line with the suggestion of the agent are presented in Figure~\ref{fig:disagreement}. We see that subjects do not find the explanations of the agent to provide convincing arguments. This shows that there is some inherent bias, and that subjects are implicitly evaluating the quality of the suggestion too, and not just the explanations. However, we notice that explanations containing Level 2 information are still seen as convincing in 40\% of the cases, compared to 21.6\% for explanations containing Level 1 information.

To control for statistical significance we perform the Kruskal-Wallis test, a non-parametric version of ANOVA which can be applied to non-normally distributed data like in our case. Results showed that there is significant difference between the condition means for each of the measurements ($p<0.001$). To control for differences between the pairwise conditions, we perform Dunn's test. Results show that the evaluation of both level 1 and level 2 explanations are significantly different from the explanations of the control group across all measurements ($p<0.01$). However, when comparing the evaluations of level 1 explanations to those of level 2 explanations, the difference is not statistically significant for any of the measurements ($p>0.05$).

This experiment allows us to answer \textbf{RQ3}: Approximately 70\% of the subjects find the explanations based on Level 1 or Level 2 information to be complete, satisfying, in line with the way the subjects reason, likely to convince the user, as well as containing sufficient information. This makes such information a good candidate for forming explanations in personal assistant agents.

\subsubsection{Experiment 2.2}\label{sec:study2_results2.2}
The goal of Experiment 2.2 was to evaluate \textbf{RQ4}. Results are presented in Table~\ref{tab:comparative_results}. First of all, for each measurement we report the answer distributions across the different scenario pairs based on which psychological characteristic was salient in the pairs. The results show that the preferences of the subjects vary between situation types. However, we notice consistency within types: for a specific pair, subjects tend to prefer the same explanation across all measurements. Given this, for simplicity we will abuse terminology and say that subjects prefer one explanation over the other in a pair of scenarios when the subjects prefer that explanation for at least four measurements. 

From the answer distributions, we notice that in situations where duty is the salient feature, subjects prefer explanations involving Level 2 information. On the other hand, in situations where negativity is the salient feature, subjects strongly prefer explanations involving Level 1 information. This seems to suggest that subjects do not like explanations that have a negative framing\footnote{The explanation involving Level 1 information was \textit{``Alice should attend Meeting 2, since in it she is meeting someone with whom she has a better relationship, and meeting with people with whom one has a better relationship are usually prioritized."}, while the explanation involving Level 2 information was \textit{``Alice should attend Meeting 2, since Meeting 1 could entail a high level of stress, and meetings that entail a low level of stress are usually prioritized."}}. For situations where the salient feature is intellect or positivity we cannot reach a clear conclusion regarding which explanation is preferred, since the results are different across pairs and seem to be context dependent.

% Please add the following required packages to your document preamble:
% \usepackage{multirow}
\begin{table}[!h]
\caption{Significance test could not be performed for the measurement `provides convincing arguments' since only a small portion of subjects made choices different from the ones of the agent and was presented with that measurement.}
\resizebox{\textwidth}{!}{
\begin{tabular}{cccccccc}
\hline
                                                                                                            & \begin{tabular}[c]{@{}c@{}}Preferred\\ Explanation\end{tabular} & Satisfying                & \begin{tabular}[c]{@{}c@{}}Sufficient\\ information\end{tabular} & Complete                  & \begin{tabular}[c]{@{}c@{}}In line with\\ user\end{tabular} & \begin{tabular}[c]{@{}c@{}}Likely to\\ convince\end{tabular} & \begin{tabular}[c]{@{}c@{}}Convincing\\ arguments\end{tabular} \\ \hline
\multirow{3}{*}{\begin{tabular}[c]{@{}c@{}}Duty-salient \\ situations\end{tabular}}                         & Level 1                                                         & 36.2\%                    & 22.9\%                                                           & 31.3\%                    & 38.6\%                                                      & 30.1\%                                                       & 40.1\%                                                         \\
                                                                                                            & Neutral                                                         & 7.2\%                     & 19.3\%                                                           & 15.7\%                    & 12\%                                                        & 12.1\%                                                       & 18.2\%                                                         \\
                                                                                                            & Level 2                                                         & 56.6\%                    & 57.8\%                                                           & 53\%                      & 49.4\%                                                      & 57.8\%                                                       & 40.1\%                                                         \\ \hline
\multirow{3}{*}{\begin{tabular}[c]{@{}c@{}}Intellect-salient \\ situations\end{tabular}}                    & Level 1                                                         & 43.4\%                    & 34.9\%                                                           & 44.6\%                    & 45.7\%                                                      & 47\%                                                         & 50\%                                                           \\
                                                                                                            & Neutral                                                         & 20.5\%                    & 26.5\%                                                           & 24.1\%                    & 14.5\%                                                      & 12\%                                                         & 25\%                                                           \\
                                                                                                            & Level 2                                                         & 36.1\%                    & 38.6\%                                                           & 31.3\%                    & 39.8\%                                                      & 41\%                                                         & 25\%                                                           \\ \hline
\multirow{3}{*}{\begin{tabular}[c]{@{}c@{}}Negativity-salient \\ situations\end{tabular}}                   & Level 1                                                         & 59\%                      & 57.8\%                                                           & 59\%                      & 53\%                                                        & 57.8\%                                                       & 50.6\%                                                         \\
                                                                                                            & Neutral                                                         & 14.5\%                    & 19.3\%                                                           & 22.9\%                    & 14.5\%                                                      & 14.5\%                                                       & 20.3\%                                                         \\
                                                                                                            & Level 2                                                         & 26.5\%                    & 22.9\%                                                           & 18.1\%                    & 32.5\%                                                      & 27.7\%                                                       & 29.1\%                                                         \\ \hline
\multirow{3}{*}{\begin{tabular}[c]{@{}c@{}}Positivity-salient \\ situations\end{tabular}}                   & Level 1                                                         & 37.3\%                    & 37.3\%                                                           & 37.3\%                    & 43.3\%                                                      & 42.4\%                                                       & 18.8\%                                                         \\
                                                                                                            & Neutral                                                         & 14.5\%                    & 28.9\%                                                           & 26.5\%                    & 15.7\%                                                      & 9.6\%                                                        & 18.8\%                                                         \\
                                                                                                            & Level 2                                                         & 48.2\%                    & 33.4\%                                                           & 36.2\%                    & 41\%                                                        & 48.2\%                                                       & 62.4\%                                                         \\ \hline
\multirow{3}{*}{Friedman's test}                                                                            & $\chi^2$                                                             & 19.935                    & 26.417                                                           & 21.549                    & 4.9594                                                      & 19.094                                                       & -                                                              \\
                                                                                                            & df                                                              & 3                         & 3                                                                & 3                         & 3                                                           & 3                                                            & -                                                              \\
                                                                                                            & p-value                                                         & \textbf{\textless{}0.001} & \textbf{\textless{}0.001}                                        & \textbf{\textless{}0.001} & 0.17                                                        & \textbf{\textless{}0.001}                                    & -                                                              \\ \hline
\multirow{6}{*}{\begin{tabular}[c]{@{}c@{}}Post-hoc analysis\\ Conover's test\\ \\ (p-values)\end{tabular}} & Duty-Intellect                                                  & 0.18                      & 0.067                                                            & 0.07                      & 0.91                                                        & \textbf{0.02}                                                & -                                                              \\
                                                                                                            & Duty-Negativity                                                 & \textbf{\textless{}0.001} & \textbf{\textless{}0.001}                                        & \textbf{\textless{}0.001} & 0.18                                                        & \textbf{\textless{}0.001}                                    & -                                                              \\
                                                                                                            & Duty-Positivity                                                 & 1.00                      & \textbf{0.02}                                                    & 0.71                      & 1.00                                                        & 0.626                                                        & -                                                              \\
                                                                                                            & Intellect-Negativity                                            & 0.488                     & 0.067                                                            & 0.29                      & 1.00                                                        & 1.00                                                         & -                                                              \\
                                                                                                            & Intellect-Positivity                                            & 0.393                     & 1.00                                                             & 1.00                      & 1.00                                                        & 1.00                                                         & -                                                              \\
                                                                                                            & Negativity-Positivity                                           & \textbf{\textless{}0.01}  & 0.199                                                            & \textbf{0.02}             & 1.00                                                        & 0.068                                                        & -                                                              \\ \hline
\end{tabular}}
\label{tab:comparative_results}
\end{table}

To control for statistical significance we perform Friedman's test, a nonparametric alternative to repeated measures ANOVA since our data is measured on an ordinal scale rather than continuous. For each measurement, the test controls whether the answers in each situation type (Duty-salient, Intellect-salient, Negativity-salient and Positivity-salient) differ. Results show that the answer distributions significantly differ ($p<0.05$) for all measurements apart from `in line with subject'. Friedman test is an omnibus test statistic, which indicates that there are significant differences in which explanations are seen as more satisfying, complete, having more sufficient information and likely to convince the user based on situation type, but does not tell which specific situation types have a significant effect on these measurements. For this, we conduct a post-hoc analysis in which we performed the Conover's test for pairwise comparisons in situation types. Confirming the insights from the answer distributions, we notice that the preferred explanations in situations where Duty is the salient feature significantly differ from situations in which Negativity is the salient feature. For the other situation types there is no significant effect across measurements.

This experiment gives some insights towards answering \textbf{RQ4}. It shows that subjects prefer explanations involving Level 2 information when duty is the salient feature, and explanations involving Level 1 information when negativity is the salient feature. However, this experiment also shows that more research is needed to determine which type of explanation is preferred for each situation. Overall, an agent that can give explanations including information from either level is beneficial, since the preferred explanation is context dependent and can vary.

\subsubsection{Open question analysis}\label{sec_study2_open_question}
\begin{figure}[!h]
    \centering
    \includegraphics{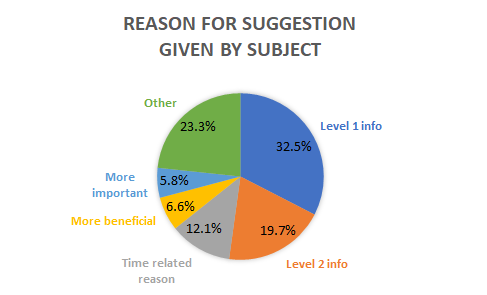}
    \caption{Distribution of reasons given by the subjects when asked why they would suggest attending a specific meeting.}
    \label{fig:open_question}
\end{figure}

After answering which meeting they would suggest to Alice, subjects were also asked about the reasons behind this suggestion. This was done to assess the type of information that users would include in their reasoning, and how it compares to the explanations given by the agent. The results are presented in Figure~\ref{fig:open_question}. The answers were analyzed by the first author in a two step procedure, following guidelines from \cite{hsieh2005three}. The first step involved summative content analysis. In it, each open answer was labeled to refer to Level 1 information, Level 2 information, or neither. To assign a label, keywords for Level 1 information were extracted from the social situation features, whereas keywords for Level 2 were extracted from the descriptors of the psychological characteristics of situations. The second step involved the open answers which did not fall under Level 1 or Level 2 information. For these answers, we performed conventional content analysis. This involves coming up with categories based on the data, rather than using preconceived categories. After reading the answers multiple times, keywords were highlighted as labels, and then clustered in cases when the keywords are logically connected. This analysis is exploratory and does not intend to provide comprehensive answers on the reasons that users have for deciding between meetings.

The results show that in more than half of the cases, subjects offered a reason that involved either the Level 1 or the Level 2 relevant feature for that pair. This confirms that subjects also reason themselves in terms of this information in many cases. Level 1 information was mentioned significantly more than Level 2 information, but this was to be expected since Level 1 information is directly present in the description of the meetings, so it is more salient. 

From this open question we can also extract other types of information that users find relevant. For instance, in 12\% of the cases subjects gave a reason that was related to temporal aspects, such as `Meeting 1 is more urgent', or `Meeting 2 is more difficult to reschedule'. This feature should be considered for inclusion to the list of Level 1 situation features, since it was consistently mentioned by subjects. Two other reasons that were consistently mentioned were `more beneficial' and `more important'. Subjects also mentioned various other similarly vague answers (e.g. `better') which did not appear consistently, therefore were clustered under `other'. Such answers show that subjects often do not explicitly dig deeper into the reasons, but offer only superficial ones.

When taking a closer look at subjects who in the open question used Level 1 or Level 2 information, we notice that the reasons that the subjects give do not necessarily match with their preferred explanations. In 43\% of the cases, in the open question subjects gave as a reason for their suggestion information from one of the levels, and in the questionnaire they preferred the explanation that included information from the other level. For instance, in the open question for Pair 5 one of the subjects says \textit{``Meeting two will be more enjoyable and less stressful"}, which fits almost perfectly with the explanation given by the agent that involves Level 2 information. However, in the questionnaire this subject always prefers significantly more the explanation that includes Level 1 information. This `flip' happens in both directions: in 50\% of cases it's from Level 1 to Level 2 and in 50\% the other way around. This suggests that there are users that want to hear explanations that differ from the reasons that they thought about themselves, providing another perspective on which explanations the agent should provide to the user.

\section{Conclusions}\label{sec:conclusions}

\subsection{Discussion}
In this work, we explore the effect of incorporating information about the psychological characteristics of situations in a socially aware agenda management support agent. To assess the benefits of this approach, we evaluate its contributions in improving the accuracy of the agent predictions, as well as in providing more satisfying explanations for the suggestions to the user. 

Automatic agenda management has been previously used as a test bed for studying how to model social relationships in agents. For instance, \cite{rist2008applying} introduce an agent that negotiates meetings on behalf of the user. The agent incorporates in its negotiation process information regarding how important the meeting is for the user, as well as information regarding the relationship quality between the user and the other person. Such an agent would benefit from the ability to automatically assess the priority of the different meetings from the point of view of the user. We hypothesized that the priority of meetings can be accurately predicted using as input the psychological characteristics of the meeting. Results in Section~\ref{sec:study1} show that psychological characteristics of situations are a significantly better predictor of the priority of situations than social situation features, thus supporting our hypothesis. Thus, using our approach for predicting the priority of social situations would be beneficial for support agents. Asking the user about the psychological characteristics of each individual situation would be a cumbersome task. For this reason, we explore whether this information can be assessed automatically. We show that using a random forest model that take as input the social situation features of a situation allows us to accurately predict the psychological characteristics of that situation. Collecting social situation features is a less invasive task, since information about social relationships can be collected once and used across multiple interactions. \cite{murukannaiah2011platys} show that active learning can be used to collect information in a less invasive manner.

In Section~\ref{sec:study2}, we show that people find explanations based on social situation features and psychological characteristics of situations to be satisfying, containing sufficient information, complete, in line with how they think, and convincing. Using brief explanations focusing on why a certain suggestion was made as opposed to the alternative led to satisfying explanations, in line with findings from related work \citep{miller2019explanation,ribera2019can}. Furthermore, we notice that when the suggestions of the agent are not in line with people's expected suggestions, they do not like the explanations. This is in line with findings reported by \cite{riveiro2021s}. Work on explanations for recommender systems \citep{tintarev2015explaining} suggests that the type of information contained the explanation affects the perceived quality of the explanation. Our work represents a first attempt in evaluating what type of information is preferred in recommendations regarding social situations. Our findings show that people prefer explanations based on psychological characteristics in situations where the level of duty is relevant, and explanations based on social situation features in situations where the level of negativity is relevant. Both types of explanations were evaluated positively, indicating that it may be beneficial if support agents were able to give explanations based on both types of information.

Overall, our results suggest that incorporating information about psychological characteristics of the user's situation can be beneficial for support agents, since it would enable them to more accurately predict information that can be used as a basis for suggestions and for explaining the suggestions to the user.

\subsection{Ethical impact}
Several ethical considerations have to be made before deploying an agent to offer support in the real world. First of all, the agent's assessments of the priority of situations can be inaccurate, thus offering to the user suggestions that can have social repercussions. For this reason, in our use case the decision remains in the hands of the user, and the agent also offers explanations for its suggestions. However, this also does not fully mitigate ethical risks. For instance, the agent might wrongly infer that a specific social situation has a high level of negativity, and inform the user about it in an explanation. However, if this is a situation which is sensitive for the user, the explanation can cause distress. Therefore, it is important to increase prediction accuracy, as well as to have more studies that assess the effects on a user of using such an agent on a daily basis.

\subsection{Limitations and Future Work}
In this work, results were based on the use case of a socially aware personal assistant agent. Future work should extend the findings for different types of support agents and other support domains. Here it will be particularly interesting to investigate if the general nature of psychological characteristics makes them a good candidate to predict other aspects of social situations besides their priority. Assuming a support agent that can assist in various tasks and different daily situations, having a common conceptual grounding for assessing the meaning of situations for the user could have advantages for human-machine meaning. Furthermore, in this paper we used a hypothetical setting in order to be able to gather larger amounts of data in a controlled way. Based on the results from this hypothetical setting, it is important to build a prototype support agent in order to test the methods in real tasks. 

While answering Research Questions 1 and 2 we found that predicting the psychological characteristics of situations accurately is crucial in order to better predict the priority of situations. In future work, we will explore other techniques, such as using natural language processing techniques to extract the psychological characteristics of situations from textual descriptions of situations. Lastly, Study 2 shows that while both social situation features and psychological characteristics of situations can be the basis of explanations given by support agents, more research is needed to determine which type of explanation to give in which situation. 

\section*{Acknowledgements}
This work is part of the research programme CoreSAEP, with project number 639.022.416, which is financed by the Netherlands Organisation for Scientific Research (NWO). This work was partly funded by the \hyperlink{https://hybrid-intelligence-centre.nl}{Hybrid Intelligence Center}, a 10-year programme funded the Dutch Ministry of Education, Culture and Science through NWO grant number 024.004.022 and by EU H2020 ICT48 project ``Humane AI Net'' under contract $\#$952026.

\bibliographystyle{plainnat}
\bibliography{references}  %%% Uncomment this line and comment out the ``thebibliography'' section below to use the external .bib file (using bibtex) .

\begin{thebibliography}{55}
\providecommand{\natexlab}[1]{#1}
\providecommand{\url}[1]{\texttt{#1}}
\expandafter\ifx\csname urlstyle\endcsname\relax
  \providecommand{\doi}[1]{doi: #1}\else
  \providecommand{\doi}{doi: \begingroup \urlstyle{rm}\Url}\fi

\bibitem[Ajmeri et~al.(2017)Ajmeri, Murukannaiah, Guo, and
  Singh]{ajmeri2017arnor}
Nirav Ajmeri, Pradeep~K Murukannaiah, Hui Guo, and Munindar~P Singh.
\newblock Arnor: Modeling social intelligence via norms to engineer
  privacy-aware personal agents.
\newblock In \emph{Proceedings of the 16th Conference on Autonomous Agents and
  MultiAgent Systems}, pages 230--238, 2017.

\bibitem[Brown et~al.(2015)Brown, Neel, and Sherman]{brown2015measuring}
Nicolas~A Brown, Rebecca Neel, and Ryne~A Sherman.
\newblock Measuring the evolutionarily important goals of situations:
  Situational affordances for adaptive problems.
\newblock \emph{Evolutionary Psychology}, 13\penalty0 (3):\penalty0 1--15,
  2015.

\bibitem[Chromik and Schuessler(2020)]{chromik2020taxonomy}
Michael Chromik and Martin Schuessler.
\newblock A taxonomy for human subject evaluation of black-box explanations in
  xai.
\newblock In \emph{ExSS-ATEC@ IUI}, 2020.

\bibitem[Cranefield et~al.(2017)Cranefield, Winikoff, Dignum, and
  Dignum]{cranefield2017no}
Stephen Cranefield, Michael Winikoff, Virginia Dignum, and Frank Dignum.
\newblock No pizza for you: Value-based plan selection in bdi agents.
\newblock In \emph{IJCAI}, pages 178--184, 2017.

\bibitem[Davison et~al.(2021)Davison, Wijnen, Charisi, van~der Meij, Reidsma,
  and Evers]{davison2021words}
Daniel~P Davison, Frances~M Wijnen, Vicky Charisi, Jan van~der Meij, Dennis
  Reidsma, and Vanessa Evers.
\newblock Words of {E}ncouragement: {H}ow {P}raise {D}elivered by a {S}ocial
  {R}obot {C}hanges {C}hildren’s {M}indset for {L}earning.
\newblock \emph{Journal on multimodal user interfaces}, 15\penalty0
  (1):\penalty0 61--76, 2021.

\bibitem[Dignum(2004)]{dignum2004model}
Virginia Dignum.
\newblock \emph{A {M}odel for {O}rganizational {I}nteraction: {B}ased on
  {A}gents, {F}ounded in {L}ogic}.
\newblock SIKS PhD Dissertation Series, 2004.

\bibitem[Dignum and Dignum(2014)]{dignum2014contextualized}
Virginia Dignum and Frank Dignum.
\newblock Contextualized planning using social practices.
\newblock In \emph{International workshop on coordination, organizations,
  institutions, and norms in agent systems}, pages 36--52. Springer, 2014.

\bibitem[Doshi-Velez and Kim(2017)]{doshi2017towards}
Finale Doshi-Velez and Been Kim.
\newblock Towards a rigorous science of interpretable machine learning.
\newblock \emph{arXiv preprint arXiv:1702.08608}, 2017.

\bibitem[Edwards and Templeton(2005)]{edwards2005structure}
John~A Edwards and Angela Templeton.
\newblock The structure of perceived qualities of situations.
\newblock \emph{European journal of social psychology}, 35\penalty0
  (6):\penalty0 705--723, 2005.

\bibitem[Elgarf et~al.(2022)Elgarf, Calvo-Barajas, Alves-Oliveira, Perugia,
  Castellano, Peters, and Paiva]{elgarf2022and}
Maha Elgarf, Natalia Calvo-Barajas, Patricia Alves-Oliveira, Giulia Perugia,
  Ginevra Castellano, Christopher Peters, and Ana Paiva.
\newblock ``and {T}hen {W}hat {H}appens?'' {P}romoting {C}hildren's {V}erbal
  {C}reativity {U}sing a {R}obot.
\newblock In \emph{Proceedings of the 2022 ACM/IEEE International Conference on
  Human-Robot Interaction}, pages 71--79. ACM, 2022.

\bibitem[Endsley(1995)]{endsley1995toward}
Mica~R Endsley.
\newblock Toward a theory of situation awareness in dynamic systems.
\newblock \emph{Human factors}, 37\penalty0 (1):\penalty0 32--64, 1995.

\bibitem[Fitzpatrick et~al.(2017)Fitzpatrick, Darcy, and
  Vierhile]{fitzpatrick2017delivering}
Kathleen~Kara Fitzpatrick, Alison Darcy, and Molly Vierhile.
\newblock Delivering cognitive behavior therapy to young adults with symptoms
  of depression and anxiety using a fully automated conversational agent
  (woebot): a randomized controlled trial.
\newblock \emph{JMIR mental health}, 4\penalty0 (2):\penalty0 e7785, 2017.

\bibitem[Fornara et~al.(2007)Fornara, Vigan{\`o}, and
  Colombetti]{fornara2007agent}
Nicoletta Fornara, Francesco Vigan{\`o}, and Marco Colombetti.
\newblock Agent {C}ommunication and {A}rtificial {I}nstitutions.
\newblock \emph{Autonomous Agents and Multi-Agent Systems}, 14\penalty0
  (2):\penalty0 121--142, 2007.

\bibitem[Gerpott et~al.(2018)Gerpott, Balliet, Columbus, Molho, and
  de~Vries]{gerpott2018people}
Fabiola~H Gerpott, Daniel Balliet, Simon Columbus, Catherine Molho, and
  Reinout~E de~Vries.
\newblock How do people think about interdependence? a multidimensional model
  of subjective outcome interdependence.
\newblock \emph{Journal of Personality and Social Psychology}, 115\penalty0
  (4):\penalty0 716, 2018.

\bibitem[Goodman and Flaxman(2017)]{goodman2017european}
Bryce Goodman and Seth Flaxman.
\newblock European union regulations on algorithmic decision-making and a
  “right to explanation”.
\newblock \emph{AI magazine}, 38\penalty0 (3):\penalty0 50--57, 2017.

\bibitem[Grosz(2012)]{grosz2012question}
Barbara Grosz.
\newblock What {Q}uestion {W}ould {T}uring {P}ose {T}oday?
\newblock \emph{AI magazine}, 33\penalty0 (4):\penalty0 73--73, 2012.

\bibitem[Gu et~al.(2018)Gu, Kelly, and Xiu]{gu2018empirical}
Shihao Gu, Bryan Kelly, and Dacheng Xiu.
\newblock Empirical asset pricing via machine learning.
\newblock Technical report, National Bureau of Economic Research, 2018.

\bibitem[Guidotti et~al.(2018)Guidotti, Monreale, Ruggieri, Turini, Giannotti,
  and Pedreschi]{guidotti2018survey}
Riccardo Guidotti, Anna Monreale, Salvatore Ruggieri, Franco Turini, Fosca
  Giannotti, and Dino Pedreschi.
\newblock A survey of methods for explaining black box models.
\newblock \emph{ACM computing surveys (CSUR)}, 51\penalty0 (5):\penalty0 1--42,
  2018.

\bibitem[Gunning and Aha(2019)]{gunning2019darpa}
David Gunning and David Aha.
\newblock Darpa’s explainable artificial intelligence (xai) program.
\newblock \emph{AI magazine}, 40\penalty0 (2):\penalty0 44--58, 2019.

\bibitem[Hoffman et~al.(2018)Hoffman, Mueller, Klein, and
  Litman]{hoffman2018metrics}
Robert~R Hoffman, Shane~T Mueller, Gary Klein, and Jordan Litman.
\newblock Metrics for explainable ai: Challenges and prospects.
\newblock \emph{arXiv preprint arXiv:1812.04608}, 2018.

\bibitem[Hsieh and Shannon(2005)]{hsieh2005three}
Hsiu-Fang Hsieh and Sarah~E Shannon.
\newblock Three approaches to qualitative content analysis.
\newblock \emph{Qualitative health research}, 15\penalty0 (9):\penalty0
  1277--1288, 2005.

\bibitem[Jameson et~al.(2014)Jameson, Berendt, Gabrielli, Cena, Gena, Vernero,
  and Reinecke]{jameson2014choice}
Anthony Jameson, Bettina Berendt, Silvia Gabrielli, Federica Cena, Cristina
  Gena, Fabiana Vernero, and Katharina Reinecke.
\newblock Choice architecture for human-computer interaction.
\newblock \emph{Foundations and Trends in Human-Computer Interaction},
  7\penalty0 (1--2):\penalty0 1--235, 2014.

\bibitem[Kepuska and Bohouta(2018)]{kepuska2018next}
Veton Kepuska and Gamal Bohouta.
\newblock Next-generation of virtual personal assistants (microsoft cortana,
  apple siri, amazon alexa and google home).
\newblock In \emph{2018 IEEE 8th Annual Computing and Communication Workshop
  and Conf.}, pages 99--103. IEEE, 2018.

\bibitem[Kola et~al.(2019)Kola, Jonker, and van Riemsdijk]{kola2019s}
Ilir Kola, Catholijn~M Jonker, and M~Birna van Riemsdijk.
\newblock Who's that? {S}ocial situation awareness for behaviour support
  agents.
\newblock In \emph{International Workshop on Engineering Multi-Agent Systems},
  pages 127--151. Springer, 2019.

\bibitem[Kola et~al.(2020{\natexlab{a}})Kola, Jonker, Tielman, and van
  Riemsdijk]{kola2020grouping}
Ilir Kola, Catholijn~M Jonker, Myrthe~L Tielman, and M~Birna van Riemsdijk.
\newblock Grouping situations based on their psychological characteristics
  gives insight into personal values.
\newblock In \emph{11th International Workshop Modelling and Reasoning in
  Context}, pages 17--26, 2020{\natexlab{a}}.

\bibitem[Kola et~al.(2020{\natexlab{b}})Kola, Tielman, Jonker, and van
  Riemsdijk]{kola2020predicting}
Ilir Kola, Myrthe~L Tielman, Catholijn~M Jonker, and M~Birna van Riemsdijk.
\newblock Predicting the priority of social situations for personal assistant
  agents.
\newblock In \emph{International Conference on Principles and Practice of
  Multi-Agent Systems}. Springer, 2020{\natexlab{b}}.

\bibitem[Kola et~al.(2022)Kola, Murukannaiah, Jonker, and
  Van~Riemsdijk]{kola2021ssa}
Ilir Kola, Pradeep~K Murukannaiah, Catholijn~M Jonker, and M~Birna
  Van~Riemsdijk.
\newblock Towards social situation awareness in support agents.
\newblock \emph{IEEE Intelligent Systems}, 2022.

\bibitem[Kop et~al.(2014)Kop, Hoogendoorn, and Klein]{kop2014personalized}
Reinier Kop, Mark Hoogendoorn, and Michel~CA Klein.
\newblock A personalized support agent for depressed patients: Forecasting
  patient behavior using a mood and coping model.
\newblock In \emph{2014 IEEE/WIC/ACM International Joint Conferences on Web
  Intelligence (WI) and Intelligent Agent Technologies (IAT)}, volume~3, pages
  302--309. IEEE, 2014.

\bibitem[Lewin(1939)]{lewin1939field}
Kurt Lewin.
\newblock Field theory and experiment in social psychology: Concepts and
  methods.
\newblock \emph{American journal of sociology}, 44\penalty0 (6):\penalty0
  868--896, 1939.

\bibitem[Lim et~al.(2009)Lim, Dey, and Avrahami]{lim2009and}
Brian~Y Lim, Anind~K Dey, and Daniel Avrahami.
\newblock Why and why not explanations improve the intelligibility of
  context-aware intelligent systems.
\newblock In \emph{Proceedings of the SIGCHI conference on human factors in
  computing systems}, pages 2119--2128, 2009.

\bibitem[Lundberg and Lee(2017)]{lundberg2017unified}
Scott~M Lundberg and Su-In Lee.
\newblock A unified approach to interpreting model predictions.
\newblock In \emph{Advances in neural information processing systems}, pages
  4765--4774, 2017.

\bibitem[Maestro-Prieto et~al.(2020)Maestro-Prieto, Rodr{\'\i}guez, Casado, and
  Corchado]{maestro2020agent}
Jose~Alberto Maestro-Prieto, Sara Rodr{\'\i}guez, Roberto Casado, and
  Juan~Manuel Corchado.
\newblock Agent {O}rganisations: {F}rom {I}ndependent {A}gents to {V}irtual
  {O}rganisations and {O}ocieties of {A}gents.
\newblock \emph{Advances in Distributed Computing and Artificial Intelligence
  Journal}, 9\penalty0 (4):\penalty0 55--70, 2020.

\bibitem[Miller(2019)]{miller2019explanation}
Tim Miller.
\newblock Explanation in artificial intelligence: Insights from the social
  sciences.
\newblock \emph{Artificial Intelligence}, 267:\penalty0 1--38, 2019.

\bibitem[Mueller et~al.(2019)Mueller, Hoffman, Clancey, Emrey, and
  Klein]{mueller2019explanation}
Shane~T Mueller, Robert~R Hoffman, William Clancey, Abigail Emrey, and Gary
  Klein.
\newblock Explanation in human-ai systems: A literature meta-review, synopsis
  of key ideas and publications, and bibliography for explainable ai.
\newblock \emph{arXiv preprint arXiv:1902.01876}, 2019.

\bibitem[Murukannaiah and Singh(2011)]{murukannaiah2011platys}
Pradeep Murukannaiah and Munindar Singh.
\newblock Platys social: Relating shared places and private social circles.
\newblock \emph{IEEE Internet Computing}, 16\penalty0 (3):\penalty0 53--59,
  2011.

\bibitem[Neerincx et~al.(2018)Neerincx, van~der Waa, Kaptein, and van
  Diggelen]{neerincx2018using}
Mark~A Neerincx, Jasper van~der Waa, Frank Kaptein, and Jurriaan van Diggelen.
\newblock Using perceptual and cognitive explanations for enhanced human-agent
  team performance.
\newblock In \emph{Int. Conf. on Engineering Psychology and Cognitive
  Ergonomics}. Springer, 2018.

\bibitem[Parrigon et~al.(2017)Parrigon, Woo, Tay, and
  Wang]{parrigon2017caption}
Scott Parrigon, Sang~Eun Woo, Louis Tay, and Tong Wang.
\newblock Caption-ing the situation: A lexically-derived taxonomy of
  psychological situation characteristics.
\newblock \emph{Journal of personality and social psychology}, 112\penalty0
  (4):\penalty0 642, 2017.

\bibitem[Pinder et~al.(2018)Pinder, Vermeulen, Cowan, and
  Beale]{pinder2018digital}
Charlie Pinder, Jo~Vermeulen, Benjamin~R Cowan, and Russell Beale.
\newblock Digital behaviour change interventions to break and form habits.
\newblock \emph{ACM Transactions on Computer-Human Interaction (TOCHI)},
  25\penalty0 (3):\penalty0 15, 2018.

\bibitem[Rauthmann et~al.(2014)Rauthmann, Gallardo-Pujol, Guillaume, Todd,
  Nave, Sherman, Ziegler, Jones, and Funder]{rauthmann2014situational}
John~F Rauthmann, David Gallardo-Pujol, Esther~M Guillaume, Elysia Todd,
  Christopher~S Nave, Ryne~A Sherman, Matthias Ziegler, Ashley~Bell Jones, and
  David~C Funder.
\newblock The situational eight diamonds: A taxonomy of major dimensions of
  situation characteristics.
\newblock \emph{Journal of Personality and Social Psychology}, 107\penalty0
  (4):\penalty0 677, 2014.

\bibitem[Reckwitz(2002)]{reckwitz2002toward}
Andreas Reckwitz.
\newblock Toward a theory of social practices: A development in culturalist
  theorizing.
\newblock \emph{European journal of social theory}, 5\penalty0 (2):\penalty0
  243--263, 2002.

\bibitem[Ribera and Lapedriza(2019)]{ribera2019can}
Mireia Ribera and Agata Lapedriza.
\newblock Can we do better explanations? a proposal of user-centered
  explainable ai.
\newblock In \emph{Joint Proceedings of the ACM IUI 2019 Workshops}, volume
  2327, pages 38--45. ACM, 2019.

\bibitem[Rist and Schmitt(2008)]{rist2008applying}
Thomas Rist and Markus Schmitt.
\newblock Applying socio-psychological concepts of cognitive consistency to
  negotiation dialog scenarios with embodied conversational characters.
\newblock \emph{Animating Expressive Characters for Social Interaction}, pages
  213--234, 2008.

\bibitem[Riveiro and Thill(2021)]{riveiro2021s}
Maria Riveiro and Serge Thill.
\newblock “that's (not) the output i expected!” on the role of end user
  expectations in creating explanations of ai systems.
\newblock \emph{Artificial Intelligence}, 298:\penalty0 103507, 2021.

\bibitem[Rosenfeld and Kraus(2018)]{rosenfeld2018predicting}
Ariel Rosenfeld and Sarit Kraus.
\newblock Predicting human decision-making: From prediction to action.
\newblock \emph{Synthesis lectures on artificial intelligence and machine
  learning}, 12\penalty0 (1):\penalty0 1--150, 2018.

\bibitem[Scott et~al.(1977)Scott, Clancey, Davis, and
  Shortliffe]{scott1977explanation}
A~Carlisle Scott, William~J Clancey, Randall Davis, and Edward~H Shortliffe.
\newblock Explanation capabilities of production-based consultation systems.
\newblock Technical report, University of Stanford, 1977.

\bibitem[Tambe(2008)]{tambe2008electric}
Milind Tambe.
\newblock Electric elves: What went wrong and why.
\newblock \emph{AI magazine}, 29\penalty0 (2):\penalty0 23--23, 2008.

\bibitem[Tintarev and Masthoff(2012)]{tintarev2012evaluating}
Nava Tintarev and Judith Masthoff.
\newblock Evaluating the effectiveness of explanations for recommender systems.
\newblock \emph{User Modeling and User-Adapted Interaction}, 22\penalty0
  (4-5):\penalty0 399--439, 2012.

\bibitem[Tintarev and Masthoff(2015)]{tintarev2015explaining}
Nava Tintarev and Judith Masthoff.
\newblock Explaining recommendations: Design and evaluation.
\newblock In \emph{Recommender systems handbook}, pages 353--382. Springer,
  2015.

\bibitem[Valstar et~al.(2016)Valstar, Baur, Cafaro, Ghitulescu, Potard, Wagner,
  Andr{\'e}, Durieu, Aylett, Dermouche, et~al.]{valstar2016ask}
Michel Valstar, Tobias Baur, Angelo Cafaro, Alexandru Ghitulescu, Blaise
  Potard, Johannes Wagner, Elisabeth Andr{\'e}, Laurent Durieu, Matthew Aylett,
  Soumia Dermouche, et~al.
\newblock Ask {A}lice: {A}n {A}rtificial {R}etrieval of {I}nformation {A}gent.
\newblock In \emph{Proceedings of the 18th ACM International Conference on
  Multimodal Interaction}, pages 419--420. ACM, 2016.

\bibitem[van~der Waa et~al.(2021)van~der Waa, Nieuwburg, Cremers, and
  Neerincx]{van2021evaluating}
Jasper van~der Waa, Elisabeth Nieuwburg, Anita Cremers, and Mark Neerincx.
\newblock Evaluating xai: A comparison of rule-based and example-based
  explanations.
\newblock \emph{Artificial Intelligence}, 291:\penalty0 103404, 2021.

\bibitem[Van~Riemsdijk et~al.(2015)Van~Riemsdijk, Jonker, and
  Lesser]{van2015creating}
M.~Birna Van~Riemsdijk, Catholijn~M. Jonker, and Victor Lesser.
\newblock Creating {S}ocially {A}daptive {E}lectronic {P}artners:
  {I}nteraction, {R}easoning and {E}thical {C}hallenges.
\newblock In \emph{AAMAS}, pages 1201--1206, 2015.

\bibitem[Vargas~Quiros et~al.(2021)Vargas~Quiros, Kapcak, Hung, and
  Cabrera-Quiros]{vargas2021}
Jose~David Vargas~Quiros, Oyku Kapcak, Hayley Hung, and Laura Cabrera-Quiros.
\newblock Individual and {J}oint {B}ody {M}ovement {A}ssessed by {W}earable
  {S}ensing as a {P}redictor of {A}ttraction in {S}peed {D}ates.
\newblock \emph{IEEE Transactions on Affective Computing}, 2021.

\bibitem[Vasilyeva et~al.()Vasilyeva, Wilkenfeld, and
  Lombrozo]{vasilyeva2015goals}
Nadya Vasilyeva, Daniel~A Wilkenfeld, and Tania Lombrozo.
\newblock Goals affect the perceived quality of explanations.
\newblock In \emph{Proceedings of the 37th Annual Conference of the Cognitive
  Science Society}.

\bibitem[Wang et~al.(2019)Wang, Yang, Abdul, and Lim]{wang2019designing}
Danding Wang, Qian Yang, Ashraf Abdul, and Brian~Y Lim.
\newblock Designing theory-driven user-centric explainable ai.
\newblock In \emph{Proceedings of the 2019 CHI conference on human factors in
  computing systems}, pages 1--15, 2019.

\bibitem[Ziegler(2014)]{ziegler2014big}
M~Ziegler.
\newblock Big five inventory of personality in occupational situations.
\newblock \emph{M{\"o}dling, Austria: Schuhfried GmbH}, 2014.

\end{thebibliography}

%%% Uncomment this section and comment out the \bibliography{references} line above to use inline references.
% \begin{thebibliography}{1}

% 	\bibitem{kour2014real}
% 	George Kour and Raid Saabne.
% 	\newblock Real-time segmentation of on-line handwritten arabic script.
% 	\newblock In {\em Frontiers in Handwriting Recognition (ICFHR), 2014 14th
% 			International Conference on}, pages 417--422. IEEE, 2014.

% 	\bibitem{kour2014fast}
% 	George Kour and Raid Saabne.
% 	\newblock Fast classification of handwritten on-line arabic characters.
% 	\newblock In {\em Soft Computing and Pattern Recognition (SoCPaR), 2014 6th
% 			International Conference of}, pages 312--318. IEEE, 2014.

% 	\bibitem{hadash2018estimate}
% 	Guy Hadash, Einat Kermany, Boaz Carmeli, Ofer Lavi, George Kour, and Alon
% 	Jacovi.
% 	\newblock Estimate and replace: A novel approach to integrating deep neural
% 	networks with existing applications.
% 	\newblock {\em arXiv preprint arXiv:1804.09028}, 2018.

% \end{thebibliography}

\end{document}